# Dynamic mitigation of two-stream instability in plasma

S. Kawata[1*] and Y. J. Gu[2, 3]

[1] Graduate School of Engineering, Utsunomiya University, 321-8585 Utsunomiya, Japan.

[2] Institute of Physics of the ASCR, ELI-Beamlines, Na Slovance 2, 18221 Prague, Czech Republic.

[3] Institute of Plasma Physics of the CAS, Za Slovankou 1782/3, 18200 Prague, Czech Republic.

**Abstract**

A dynamic mitigation mechanism of the two-stream instability is discussed based on a phase control for plasma and fluid instabilities. The basic idea for the dynamic mitigation mechanism by the phase control was proposed in the paper [Phys. Plasmas 19, 024503(2012)]. The mitigation method is applied to the two-stream instability in this paper. In general, instabilities appear from the perturbations, and normally the perturbation phase is unknown. Therefore, the instability growth rate is discussed in fluids and plasmas. However, if the perturbation phase is known, the instability growth can be controlled by a superimposition of perturbations imposed actively. For instance, a perturbed driver induces a perturbation to fluids or plasmas; if the perturbation induced by the perturbed driver is oscillated actively by the driver oscillation, the perturbation phase is known and the perturbation amplitude can be controlled, like a feedforward control. The application result shown in this paper demonstrates that the dynamic mitigation mechanism works well to smooth the non-uniformities and mitigate the instabilities in plasmas.

[*] Shigeo Kawata, Prof. Dr., Graduate School of Engineering, Utsunomiya University, Yohtoh 7-1-2, Utsunomiya 321-8585, Japan. E-mail: kwt@cc.utsunomiya-u.ac.jp

# 1. Introduction

A dynamic mitigation mechanism is presented and discussed for the plasma two-stream instability in this paper. In general, instabilities grow from a perturbation, and normally the perturbation phase is unknown. Therefore, it would be difficult to control the perturbation phase, and usually the instability growth rate is discussed. However, if the perturbation phase is controlled and known, we can find a new way to control the instability growth. One of the most typical and well-known mechanisms is the feedback control in which the perturbation phase is detected and the perturbation growth is controlled or mitigated or stabilized [1]. In fluids and plasmas, it is difficult to detect the perturbation phase and amplitude. However, even in fluids and plasmas, if we can actively impose the perturbation phase by the driving energy source wobbling or oscillation, and therefore, if we know the phases of the perturbations, the perturbation growth can be controlled in a similar way as the feedforward control (see Fig. 1) [1-4]. In instabilities, one mode of the initial perturbation, from which an instability grows, may have the form of $a = a_0 e^{ikx+\gamma t}$, where $a_0$ is the amplitude, $k = 2\pi/\lambda$ is the wave number, $\lambda$ the wave length and $\gamma$ the growth rate of the instability. An example initial perturbation is shown in Fig. 1(a). At $t$=0 the perturbation is imposed. The initial perturbation may grow at instability onset. After $\Delta t$, if the feedback control works on the system, another perturbation, which has an inverse phase with the detected amplitude at $t$=0, is actively imposed (see Fig. 1(b)), and then the actual perturbation amplitude is well mitigated as shown in Fig. 1(c). This is an ideal example for the instability mitigation. This control mechanism is apparently different from the dynamic stabilization mechanism shown in the previous works in Refs. 5-9, and in Section 2 the difference is clarified.

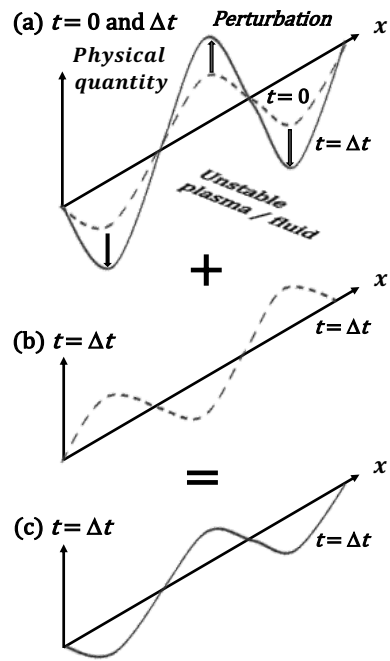

Fig. 1 An example concept of plasma / fluid control. (a) At $t=0$ a perturbation is imposed. The initial perturbation may grow at instability onset. (b) After $\Delta t$, if another perturbation, which has an inverse phase with the detected amplitude at $t=0$, is actively imposed, (c) the actual perturbation amplitude is mitigated well after the superposition of the initial and additional perturbations.

The phase control is applied to mitigate the instability growth in plasmas and fluids. For example, the growth of the filamentation instability [10-14] driven by a particle beam or jet would be controlled by the beam axis oscillation or wobbling. The oscillating and modulated beam induces the initial perturbation and also define the perturbation phase. Therefore, the successive phase-defined perturbations are superimposed, and the instability growth is mitigated. Another example can be found in heavy ion beam inertial fusion (HIF). In inertial confinement fusion (ICF) [15, 16], the fusion fuel compression is essentially important to reduce an input driver energy [17, 18], and the implosion uniformity is one of critical issues to compress the fusion fuel pellet stably [18, 18]. Therefore, the Rayleigh-Taylor Instability (RTI) stabilization or mitigation is attractive to minimize the fusion fuel mix. The heavy ion accelerator could have a capability to provide a beam axis wobbling with a high frequency [19-21]. The wobbling heavy ion beams also define the perturbation phase [21, 22]. This means that the perturbation phase is known, and so the successively imposed perturbations are superimposed on fluids and plasmas.

In this paper we first compare our dynamic mitigation mechanism with another dynamic stabilization mechanism proposed by Kapitza[5], in which the basic equation is modified by adding a strong oscillating force to create a new stable window in the system. Then we present the dynamic mitigation for the plasma two-stream instability numerically.

## 2. Dynamic mitigation mechanism of plasma instabilities under a phase control

In plasmas the perturbation phase and amplitude cannot be measured dynamically. However, by using a wobbling beam or an oscillating beam or so[19, 20], the perturbation is actively imposed from the outside of the system. In this case, the amplitude and phase of the perturbation can be defined by the input driver beam wobbling or so at least at the linear phase, like the feedforward control theory. In fluids and plasmas, it would be difficult to realize a perfect feedback control [1], however a part of it can be adapted to the instability mitigation in plasmas and fluids. Actually, heavy ion beam accelerators would provide a controlled wobbling or oscillating beam with a high frequency [19, 20].

If the driver beam wobbles uniformly in time, the imposed perturbation for a physical quantity of $F$ at $t = \tau$ may be written as

$$F = \delta F e^{i\Omega\tau} e^{\gamma(t-\tau)+i\vec{k}\cdot\vec{x}}. \tag{1}$$

Here $\delta F$ is the amplitude, $\Omega$ the wobbling or oscillation frequency, and $\Omega\tau$ the phase shift of the superimposed perturbations. At each time $t = \tau$, the wobbler provides a new perturbation with the controlled phase shifted and amplitude defined by the driving wobbler itself. After the superposition of the perturbations, the overall perturbation is described as

$$\int_0^t d\tau \; \delta F e^{i\Omega\tau} e^{\gamma(t-\tau)+i\vec{k}\cdot\vec{x}} \propto \frac{\gamma+i\Omega}{\gamma^2+\Omega^2} \delta F e^{\gamma t} e^{i\vec{k}\cdot\vec{x}}. \tag{2}$$

At each time of $t = \tau$ the driving wobbler provides a new perturbation with the shifted phase. Then, each perturbation grows with the factor of $e^{\gamma t}$. At $t > \tau$ the superimposed overall perturbation growth is modified as shown above. When $\Omega \gg \gamma$, the perturbation amplitude is reduced by the factor of $\gamma/\Omega$, compared with that for the pure instability growth ($\Omega = 0$) [2-4]. If the growth rate $\gamma$ is negative, the system concerned is stable. Even in this case, Eq. (2) shows that the non-uniformity is smoothed by the mitigation method. In the mitigation mechanism, the wobbling trajectory is under the control, for example, by a beam accelerator or so. The superimposed perturbation phase and amplitude are controlled, and the overall perturbation growth is also controlled.

From the analytical expression for the physical quantity $F$ in Eq. (2), the mechanism proposed in this paper does not work, when $\Omega \ll \gamma$. Only the modes, fulfilling the condition of $\Omega \geq \gamma$, experience the instability mitigation through the wobbling behavior. For RTI, the growth rate $\gamma$ tends to become larger for a short wavelength. If $\Omega \ll \gamma$, the modes cannot be mitigated. In addition, if there are other sources of perturbations in the physical system and if the perturbation phase and amplitude are not controlled, this dynamic mitigation mechanism also does not work. In this sense the dynamic mitigation mechanism is not almighty.

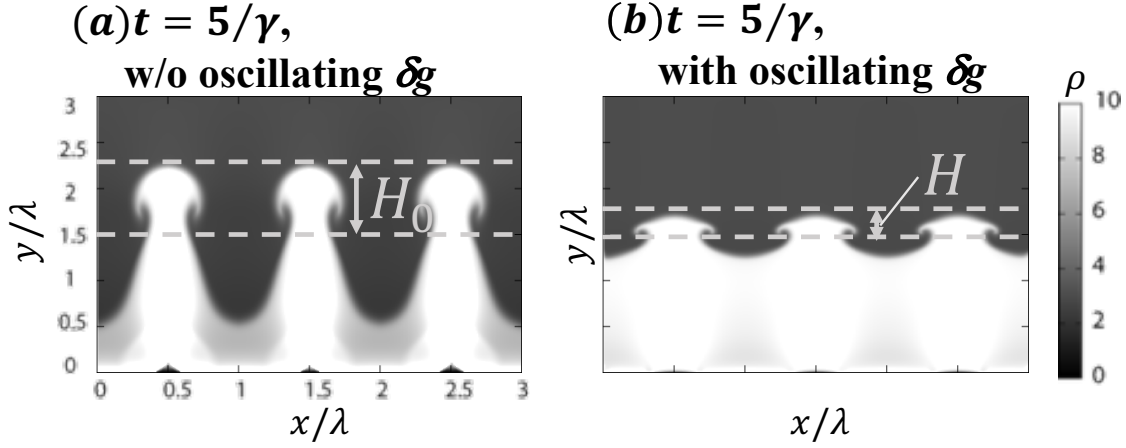

Fig. 2 Example simulation results for the Rayleigh-Taylor instability (RTI) mitigation. $\delta g$ is 10% of the acceleration $g_0$ and oscillates with the frequency of $\Omega=\gamma$. As shown above and in Eq. (2), the dynamic instability mitigation mechanism works well to mitigate the instability growth.

Figure 2 shows an example simulation for RTI, which has one mode. In this example, two stratified fluids are superimposed under an acceleration of $g = g_0 + \delta g$. The density jump ratio between the two fluids is 10/3. In this specific case the wobbling frequency $\Omega$ is $\gamma$, the amplitude of $\delta g$ is $0.1 g_0$, and the results shown in Figs. 2 are those at $t = 5/\gamma$. In Fig. 2(a) $\delta g$ is constant and drives the RTI as usual, and in Fig. 2(b) the phase of $\delta g$ oscillates with the frequency of $\Omega$ as stated above for the dynamic instability mitigation in this section. The RTI growth mitigation ratio is 72.9% in Fig. 2 at $t = 5/\gamma$. The growth mitigation ratio is defined by ($H_0$ - $H$)/$H_0$×100%. Here $H$ is defined as shown in Figs. 2, $H_0$ shows the deviation amplitude of the two-fluid interface in the case in Fig. 2(a) without the oscillation ($\Omega = 0$), and $H$ shows the deviation for the other case with the oscillation ($\Omega \neq 0$). The example simulation results support well the effect of the dynamic mitigation mechanism. Other multi modes RTI analyses are found in Ref. 23 [23].

In Refs. 6-9 one dynamic stabilization mechanism was proposed to stabilize the RTI based on the strong oscillation of acceleration. In this mechanism, the total acceleration oscillates strongly, and the additional oscillating force is added to create a new stable window in the system. Originally this dynamic stabilization mechanism was proposed by P. L. Kapitza [5], and it was applied to the stabilization of an inverted pendulum. The inverted pendulum is an unstable system, and a strongly and rapidly oscillating acceleration is applied on the system in Ref. 5, and then the inverted pendulum system creates a stable window. In this case, the equation for the unstable system is modified, and has another force term coming from the oscillating acceleration. In this mechanism, the growth rate is modified by the strongly oscillating acceleration.

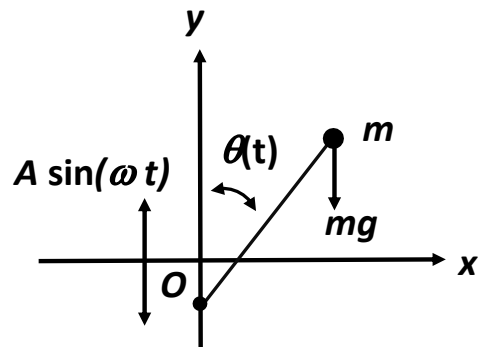

Fig. 3 Kapitza's pendulum, which is stabilized by applying an additional strong and rapidly oscillating acceleration $A \sin \omega t$.

When the inverted pendulum shown in Fig. 3 is subjected by a strongly oscillating acceleration $A \sin \omega t$, we obtain the following Mathieu-type equation [24] for $\theta(t)$:

$$\frac{d^2\theta(t)}{dt^2} = \frac{g}{l} \theta(t) - A\omega^2 \theta(t) \sin \omega t \qquad (3)$$

Here $l$ is the length of the pendulum. When $A=0$, the inverted pendulum becomes unstable. However, the second term of the righthand side is added to the system, and stable windows appear in the inverted pendulum system [5, 24]. In Eq. (3) the stability condition is described as $A - 0.5 < 2g/(l\omega^2) < A^2$.[24] The stability condition shows that the additional acceleration oscillation at the second term of the righthand side of Eq. (3) should be very fast, and the amplitude of $A$ must satisfy the stability condition shown above.

This dynamic stabilization mechanism works on the inverted pendulum in Fig. 3. However, it would be difficult to apply this mechanism to our tall buildings, bridges or large structures in our society and also to fluids and plasmas. In Ref. 25, this dynamic stabilization mechanism is applied to the two-stream instability stabilization, in which the classical two-stream instability driven by a constant relative drift velocity is modified by the additional oscillation on the relative velocity [25].

## 3. Mitigation of two-stream instability

In plasmas the two-stream instability appears commonly, when one component has a relative speed to another component [26]. In this subsection, two identical counter-streaming cold electron components are employed. The immobile background cold ions neutralize the electron charge. The number density $n_e$ of each electron beam is $10^4$ /m$^3$, and each streaming speed $v_{de}$ in the $x$ direction is $\pm 5.77 \times 10^{-3} c$. Here $c$ shows the speed of light. The electron plasma frequency $\omega_{pe}$ is about $5.64 \times 10^3$ /s. The maximal growth rate $\gamma_{max}$ of the two-stream instability is $\sim 1.41 \times 10^3$ /s at the wavelength of $\lambda_{max} \sim 22.3$ km. We employ EPOCH3D [27] to simulate the electron two-stream instability. The periodic boundary conditions are employed in all the directions in the EPOCH simulations.

Figures 4 present the two-stream instability growth. The two-stream electron beams of $n_{e+}$ and $n_{e-}$ counter each other in the $x$ direction in this example case. The subscripts of ± at $n_e$ represent the identical counter-streaming electron beams moving in +x and -x. Figure 4(a) shows the initial electron number density $n_{e+}$ of the electron beam moving in +x, Fig. 4(b) presents the electron number density $n_{e+}$ at $t$=1.4ms, and Fig. 4(c) the phase space map in the momentum $P_x$ versus $x$ space at 15km≤ $z$ ≤16km. One of the two electron beams moving in the +x direction has the initial perturbation $\delta n_{e+}$ in $n_{e+}$ with the amplitude of 5% as shown in Fig. 4(a). In the $x$ direction two modes are imposed, and one mode is introduced in $y$ and $z$ as the initial perturbation.

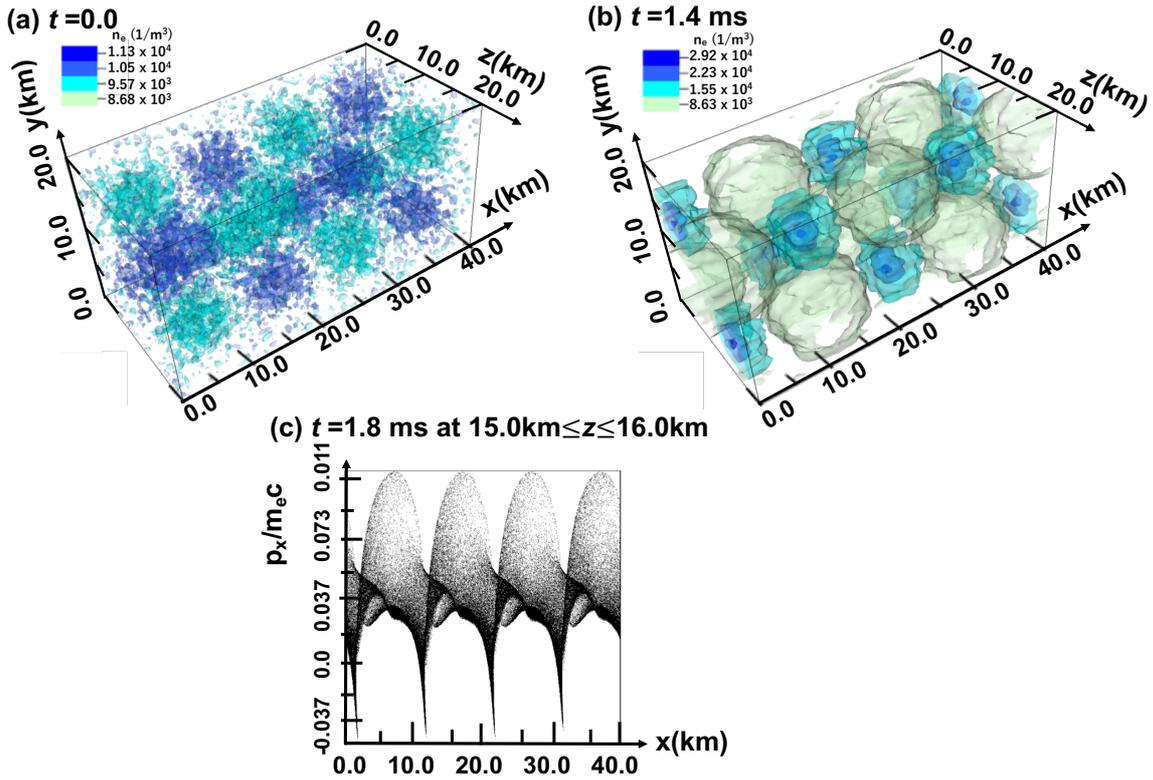

Fig. 4 Two-stream instability by two counter-streaming electron beams. The two identical electron beams counter each other. Initially (a) one of the two electron beams moving in +$x$ has the density perturbation in longitudinal and transverse. Two modes in $x$ and one mode in $y$ and $z$ are accommodated in this example case. (b)At $t$=1.4ms, the two-stream instability grows. The phase space map is also shown in (c) in $P_x$ versus $x$.

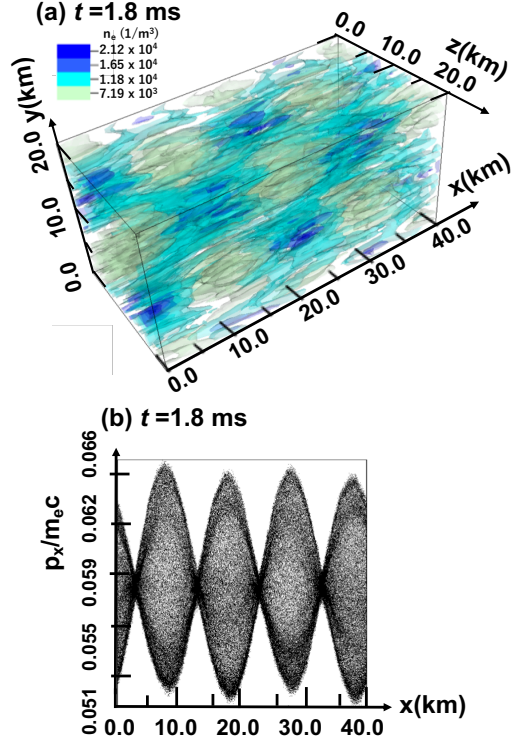

Fig. 5 Two-stream instability by two counter-streaming electron beams with the wobbling motion of $n_{e+}$. In this case, initially one of the two electron beams moving in +x has the density perturbation in $n_{e+}$ with the wobbling motion. (a)The electron number density of $n_{e+}$ is shown at $t$=1.8ms. The phase space map is also shown in (b) in $P_x$ versus $x$ at $t$=1.8ms at 15km≤ $z$ ≤16km. The two-stream instability growth is not significant compared with Figs. 4(b) and (c).

Figures 5 shows the two-stream instability by two counter-streaming electron beams with the wobbling motion of $n_{e+}$. In this case, initially one of the two electron beams moving in +x has the density perturbation in $n_{e+}$ with the wobbling motion. The wobbling motion in this case is introduced in the perturbation $\delta n_{e+}$ of the amplitude of 5% additionally with the sinusoidal oscillation in $y$ and $z$. The spatial amplitude of the wobbling oscillation is 10km, that is, the half wavelength of the $\delta n_{e+}$ perturbation in both $y$ and $z$, and the oscillation frequency is defined as that 3 oscillations are put in the one wavelength (10km) of the $\delta n_{e+}$ perturbation. Figure 5(a) shows the electron number density of $n_{e+}$ is shown at $t$=1.8ms, and Fig. 5(b) presents the phase space map in $P_x$ versus $x$ at $t$=1.8ms at 15km≤ $z$ ≤16km. The two-stream instability growth is not significant compared with Figs. 5(b) and (c).

Figure 6 shows the time sequence of the electric field energy ($\propto E_x^2$) in the $x$ direction for the two-stream instabilities with and without the wobbling motion of $n_{e+}$. At $t$=1.8ms the electric field energy is reduced by 95.4% in this specific case. The two-stream instability onset is significantly delayed by the wobbling motion of the driver electron beam. Figures 5 and 6 demonstrate that the wobbling behavior of the perturbed electron beam mitigates the two-stream instability growth successfully.

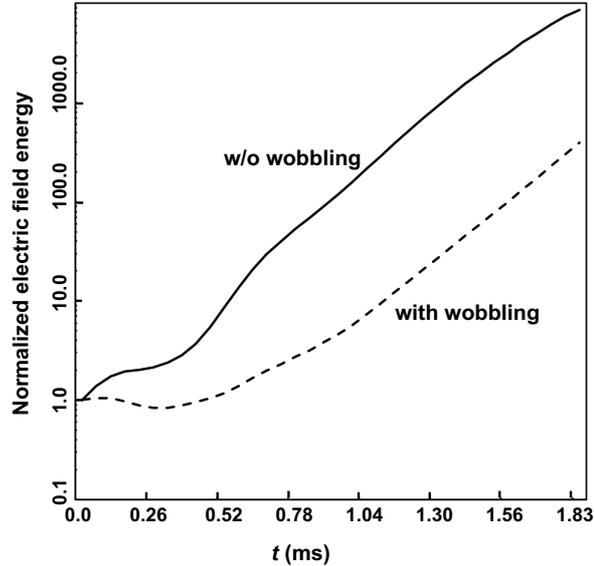

Fig. 6 The time histories of the longitudinal electric field energy associated with the two-stream instability by two counter-streaming electron beams with and without the wobbling motion of $n_{e+}$. At $t$=1.8ms the electric field energy is reduced by 95.4% in this specific case. The two-stream instability onset is significantly delayed by the wobbling motion of the driver electron beam.

## 4. Discussions and summary

In this paper we have discussed the dynamic stabilization and mitigation of the two-stream instability in plasmas. The dynamic mitigation mechanism based on the phase control was proposed and applied to the mitigation of plasma instabilities including the RTI, the filamentation instability, and also the fuel target implosion in heavy ion inertial fusion (HIF) [1-3, 21]. Originally the dynamic stabilization mechanism comes from the feedforward control, which is widely used to stabilize tall building, structures, etc. in our society. On the other hand, the dynamic stabilizations [4-6], based on the "Kapitza's pendulum" [4], introduce a new strong oscillating force into the basic equation, and then the governing equation is modified by the additional term to create a new stable window in the system. Therefore, the growth rate is modified, and the stable window appears in the system. As we discussed in this paper, we can actively apply the perturbations. Before moving to the system disruption or before developing the non-linear phase, the additional perturbations, which should have the reverse phase, are applied actively, and then the superimposed total amplitude could be mitigated by the phase control, discussed in the paper. The dynamic mitigation mechanism works to smooth the non-uniformity and delay the instability onset in fluids and plasmas.

**Acknowledgements:** This work was partly supported by Japan Society for the Promotion of Science (JSPS), Ministry of Education, Culture, Sports, Science and Technology (MEXT), Japan / U.S. Cooperation in Fusion Research and Development, Center for Optical Research and Education, Utsunomiya University (CORE), and Institute of Laser Engineering, Osaka University (ILE). This work was also partially supported by the project HiFi (CZ.02.1.01/0.0/0.0/15_003/0000449) and project ELI: Extreme Light Infrastructure (CZ.02.1.01/0.0/0.0/15_008/0000162) from European Regional Development. Computational resources were partially provided by the ECLIPSE cluster of ELI Beamlines. The Authors would like to appreciate to J. Zhang, Z. M. Sheng, S. M. Weng, S. Weber, S. Bulanov, A. Andreev and Y. Y. Ma for their fruitful discussions on this subject.